# Modeling the Thermal Behavior of Photopolymers for In-Space Fabrication

Jonathan Ericson[a], Daniel Widerker[a], Eytan Stibbe[b], Mor Elgarisi[a], Yotam Katzman[a], Omer Luria[a], Khaled Gommed[a], Alexey Razin[a], Amos A. Hari[a], Israel Gabay[a,‡], Valeri Frumkin[a,§], Hanan Abu Hamad[c], Ester Segal[c], Yaron Amouyal[d], Titus Szobody[e], Rachel Ticknor[f], Edward Balaban[f] & Moran Bercovici[a,g,*]

[a] Faculty of Mechanical Engineering, Technion - Israel Institute of Technology, Haifa, Israel

[b] The Rakia Mission, 3 Shadal St. Tel Aviv-Yafo, Israel

[c] Faculty of Biotechnology and Food Engineering, Technion - Israel Institute of Technology, Haifa, Israel

[d] Faculty of Materials Science and Engineering, Technion - Israel Institute of Technology, Haifa, Israel

[e] Department of Chemical and Biomolecular Engineering, Rice University, Houston, TX

[f] NASA Ames Research Center, Moffett Blvd., Moffett Field, CA, USA

[g] Department of Materials, ETH Zürich, Switzerland

[‡] Current affiliation: School of Chemical and Biomolecular Engineering and School of Integrative Plant Science, Cornell University, Ithaca, NY, USA

[§] Current affiliation: Department of Mechanical Engineering, Boston University, Boston, MA, USA

[*] Corresponding author: mberco@technion.ac.il

## Abstract

Future long-duration space missions will require in-situ, on-demand manufacturing of tools and components. Photopolymer-based processes are attractive for this purpose due to their low energy requirements, volume efficiency, and precise control of curing. However, photopolymerization generates significant heat, which is difficult to regulate in microgravity where natural convection is absent, leading to defects such as surface blistering and deformation. In this work, we combine experimental studies and modeling to address these thermal challenges. We report results from International Space Station (ISS) experiments and a dedicated parabolic flight campaign, which confirm that suppressed convective heat transfer in microgravity exacerbates thermal buildup and defect formation. Building on these observations, we present a predictive thermal model that couples heat transfer, light absorption, and evolving material properties to simulate polymerization and temperature evolution under terrestrial and microgravity conditions. Laboratory validation demonstrates strong agreement between model predictions and measured temperature profiles. Applying the model to the ISS experiments, we show that the model accurately reproduces experimentally observed blistering in TJ-3704A, a commercial acrylate-based polymer resin, while also predicting defect-free outcomes for Norland optical adhesives. The model functions as a design tool for defect-free in-space manufacturing, enabling selection of polymer properties, exposure strategies, and environmental conditions that together inhibit excess thermal buildup, paving the way for scalable, reliable in-situ manufacturing during future missions.

## 1. Introduction

The most significant human presence in space to date is the International Space Station (ISS), located just 400 kilometers above Earth. While 13 metric tons of spares are stored on board, the logistics model for its operation still requires more than 3 metric tons of resupply from Earth each year. It is clear that for future long-duration missions, such as crewed Mars expeditions, where resupply from Earth is unavailable, a fundamental shift in mission planning is essential. Such missions will need to be able to manufacture tools, replacement parts, and functional components in-space and on-demand[1].

Since 2011, NASA's In Space Manufacturing project (ISM) has been actively developing capabilities for on-demand in-space manufacturing and repair. Numerous approaches for in-space manufacturing have been pursued. The majority of methods are based on polymer additive-manufacturing such as FDM 3D-printing (e.g. Made In Space's 3DP and AMF projects, Tethers Unlimited Inc. Refabricator), direct robotic extrusion of photopolymers (DREPP[2]), selective UV exposure of photopolymers (VAM: SpaceCAL[3], DLP: Auxilium Biotechnologies' AMP-1[4], LCD: Photocentric's CosmicMaker[5]), and our own approach, Fluidic Shaping[6,7].

Photopolymers are particularly attractive for space as they require a very small energy budget and infrastructure in comparison to other materials. They allow precise control of polymerization initiation and location, while eliminating the need for extensive heating of feed material or for on-site homogeneous and bubble-free mixing of multi-component resins. Despite this mechanical simplification, the thermodynamics of polymerization remains a challenge. The photopolymerization reaction is inherently exothermic, releasing significant heat that, if unaccounted for, may undermine successful fabrication. At high temperatures, photopolymer resins are prone to evaporation of volatiles, surface deformations, and blistering (see Supplementary Video 3), and in extreme cases, loss of chemical stability[8]. These challenges exist in terrestrial fabrication but are exacerbated in space where temperature regulation is hindered by the absence of free convection, and even more so in large structures, where the surface area to volume ratio diminishes.

As shown in Figure 1, during our 2022 ISS experiment (see Supplementary Video 1), we observed firsthand the detrimental effects of microgravity on photopolymer based fabrication.[9] As part of the RAKIA Ax-1 mission, astronaut Eytan Stibbe demonstrated in-space manufacturing of optical lenses using our Fluidic Shaping method (Figure1A-B). Fluidic Shaping exploits the dominancy of surface-tension in the microgravity environment to naturally shape liquid polymer resins into smooth spherical lenses, which are then solidified by polymerization.[6] During the experiment, we used three different photopolymers: TJ-3704A, a commercial acrylate-based resin also sold under the brand name 'VidaRosa' (Dongguan Tianxingjian Electronic Technology Co., Ltd, Guandong, China), and NOA61 & NOA63 (Norland Optical Adhesive, Norland Products Incorporated, New Jersey, USA). In ground-based tests, we produced high-quality optics using all these materials (Figure 1D), yet to our surprise, during the ISS experiment, the lenses made from the TJ-3704A exhibited severe blistering on their free surface (Figure 1C) – a phenomenon we had not observed until then. This was further accompanied by deformations of the plastic frames and noticeable release of fumes (see Supplementary Video 2).

We hypothesize that the unexpected behavior of TJ-3704A is the result of extreme thermal build-up during polymerization, caused by the absence of natural convection in microgravity. This suggests that understanding and predicting the temperature distribution and propagation during the fabrication process is crucial to successful photopolymerization-based manufacturing in space. Such predictions are, however, not straightforward as the temperature is affected by a large number of parameters, including material properties, thermal boundary conditions, excitation light propagation, and sample geometry. Furthermore, the ability to experimentally replicate on Earth the weightless conditions under which technologies must operate in space (via parabolic flights, ballistic projectiles, and drop towers) is limited to brief periods that are usually too short to simulate realistic fabrication processes.

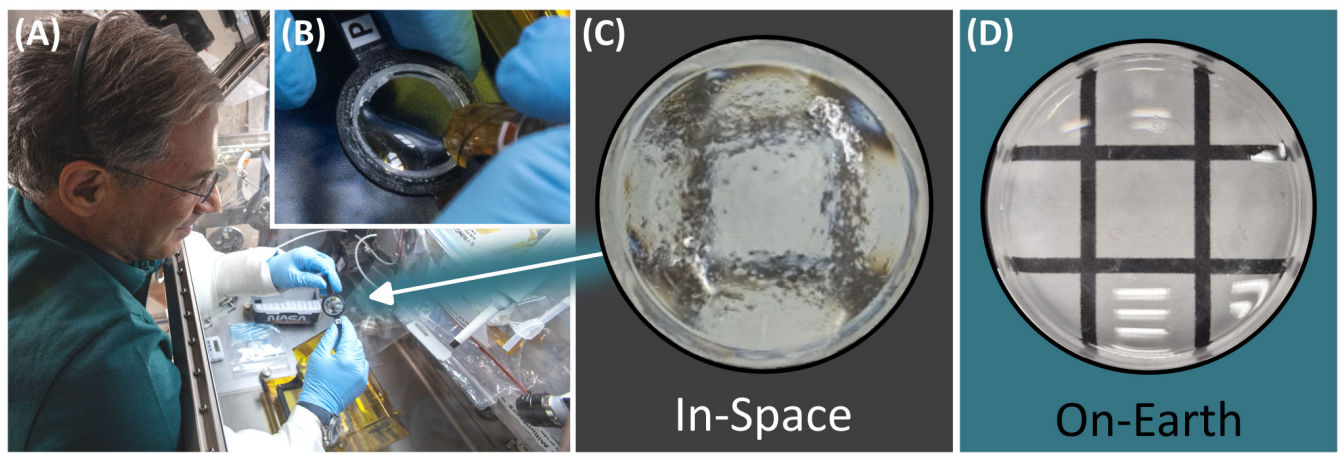

**Figure 1. Observations of heat induced defects during in-space manufacturing of photopolymer lenses.** **(A)** An image of astronaut Eytan Stibbe conducting the fluidic lens experiment aboard the International Space Station. **(B)** The astronaut injected photopolymer resins into ring-shaped frames. In microgravity, the dominant surface tension, along with boundary conditions and the injected volume, determine the lens's steady-state topography. After a short settling period, the resin was cured with UV light, resulting in a solid optical component within minutes. **(C)** During the experiment, the astronaut noticed that the lenses made of TJ-3704A exhibited severe blistering on their surface. **(D)** A typical fluidic lens fabricated on Earth: a reference lens fabricated on Earth, under neutral buoyancy conditions, shows smooth, defect-free surfaces.

In this work, we first confirmed our hypothesis using a set of parabolic flight experiments that reproduced the observations on-board the ISS under controlled conditions and clearly linked microgravity, polymer temperatures, and surface defects. Building on these results, we developed an experimentally validated model that predicts polymerization states and temperature profiles during photopolymerization-based fabrication. The model is based on the unsteady heat equation with material properties that vary in space and time as a function of local temperature and polymer phase state. Heat generation is expressed as a source term that originates from the photopolymerization reaction, and which depends on the local light intensity, modeled by Beer-Lambert absorption. The model applies to both terrestrial and microgravity environments by prescribing appropriate heat transfer boundary conditions and is applicable to any photopolymer following characterization of the polymer's thermal and optical properties.

We demonstrate the use of the model as a design tool for determining the conditions under which successful, defect-free, polymerization can be achieved in a given environment. Moreover, we show that given a specific manufacturing approach and geometry, the model can be used to determine the constraints for suitable material properties. Finally, we use the model to investigate polymer surface blistering observed during our fluidic lens fabrication on the ISS. The model reveals significant thermal buildup in microgravity due to the absence of convection and accurately captures our observations in photopolymerization of TJ-3704A and Norland optical adhesives in space.

## 2. Methods

### 2.1. Investigation of photopolymerization in microgravity

We hypothesize that the thermal defects observed in our ISS experiment can be attributed to the absence of natural convection in microgravity, which led to a significant reduction in heat dissipation. To test this hypothesis, we conducted a series of parabolic flight experiments in which photopolymer samples were UV-cured while recording their thermal response. A parabolic flight consists of repeated free-fall maneuvers providing ~15 sec intervals of microgravity, as well as 'straight and level' flight in proper acceleration of 1g. By systematically repeating the experiments in both 1g and microgravity conditions, and applying either natural or forced convection cooling, we were able to isolate the respective roles of gravity and convective heat transfer, serving as the basis for our analytical model. Additionally, these experiments allowed us to reproduce and analyze the polymer behavior previously observed in our ISS study.

We conducted the experiments aboard two consecutive parabolic flights. As shown in Figure 2, each flight carried 16 custom, airtight desiccators, each containing two to four 30 mm wide polystyrene petri dishes, each filled with photopolymer resin up to a predetermined height (1-5 mm) – dimensions that correspond to the lenses produced during our ISS experiment. To prevent cross-contamination, each desiccator housed only a single polymer formulation. We tested all three photopolymers used in the ISS experiment: TJ-3704A, NOA61, and NOA63. The outer walls of the desiccators were lined with UV-resistant Kapton tape to shield them from ambient light. Throughout the flights, we initiated the photopolymerization of the samples within the desiccators (two desiccators at a time), via remotely controlled 365 nm LEDs mounted on the desiccator lids, operated at various pre-calibrated UV intensities. We embedded thermocouples at different depths within the polymer samples to measure the temperature evolution. To regulate airflow, we used voltage-controlled fans mounted inside the chambers to produce laminar flow at approximately 4 m/s parallel to the polymer's free surface.

To ensure proper control experiments, we activated 6 of the 16 desiccators (per flight) during the 1g 'straight and level' flight segments. Turning the fans off resulted in natural convection under 1g conditions and no convection in 0g. Activating the fans established consistent forced convection in both gravity regimes. We logged the thermocouple measurements as well as the UV-LED and fan control-signals at 25 Hz, and time-synchronized them with inertial data gathered by an Arduino-based inertial measurement unit (IMU) which was mounted to the experimental setup.

We used remotely activated GoPro cameras - equipped with custom 35 mm macro-lenses and mounted inside 8 of the 16 desiccators - to capture the photopolymerization process and the emergence of thermal deformations. As shown in Figure 2C, these recordings documented visible evidence of thermal effects such as boiling, bubbling, and evaporation as well as the formation of surface blisters similar to those previously observed in our ISS experiments.

Figure 2D presents a blister density count (normalized by the largest value of all samples) as obtained from the polymerized TJ-3704A samples, as a function of the maximum temperature recorded in the sample during the experiments. The three datasets presented correspond to natural convection during the 'straight and level' flight (1g, fan off), natural convection during microgravity (0g, fan off), and forced convection during microgravity (0g, fan on). Although significant defects were observed in some samples in 1g conditions, the absence of gravity exacerbated the issue. Samples polymerized in microgravity exhibited both higher average temperatures and approximately double the blister count compared to those polymerized in 1g, under otherwise identical experimental conditions. Introducing fan-induced forced convection had a pronounced cooling effect, even at the modest airflow velocities tested. Average sample temperatures dropped by approximately 20 °C, and thermal defects were virtually non-existent in the fan-cooled samples. Interestingly, measured temperatures did not exceed ~ 105 °C regardless of experimental parameters, which leads us to suspect that substantial heat-loss through devolatilization within the polymer occurs above a critical temperature[8]. This is supported by the visual observations in the video recordings (see Supplementary Video4).

These experimental findings support our hypothesis that the effect of gravity on photopolymerization is primarily through the suppression of natural convective heat transfer. Although fan-induced airflow effectively mitigated thermal buildup in our relatively thin samples, it is not a viable solution for thicker and larger structures and is entirely impractical for construction in open space, where no atmosphere exists. In addition, the induced airflow across the liquid-air interface can distort the geometry of cured components, making fan cooling unsuitable for objects where precision is of high importance, such as optics. However, it is likely that other parameters, such as polymer thickness, UV intensity, and material properties, can be adjusted to enable defect-free fabrication in space. As the combined influence of these parameters on thermal behavior is complex and not easily predictable, we now turn to the development of a model which incorporates these parameters and is capable of accurately forecasting temperature evolution during fabrication.

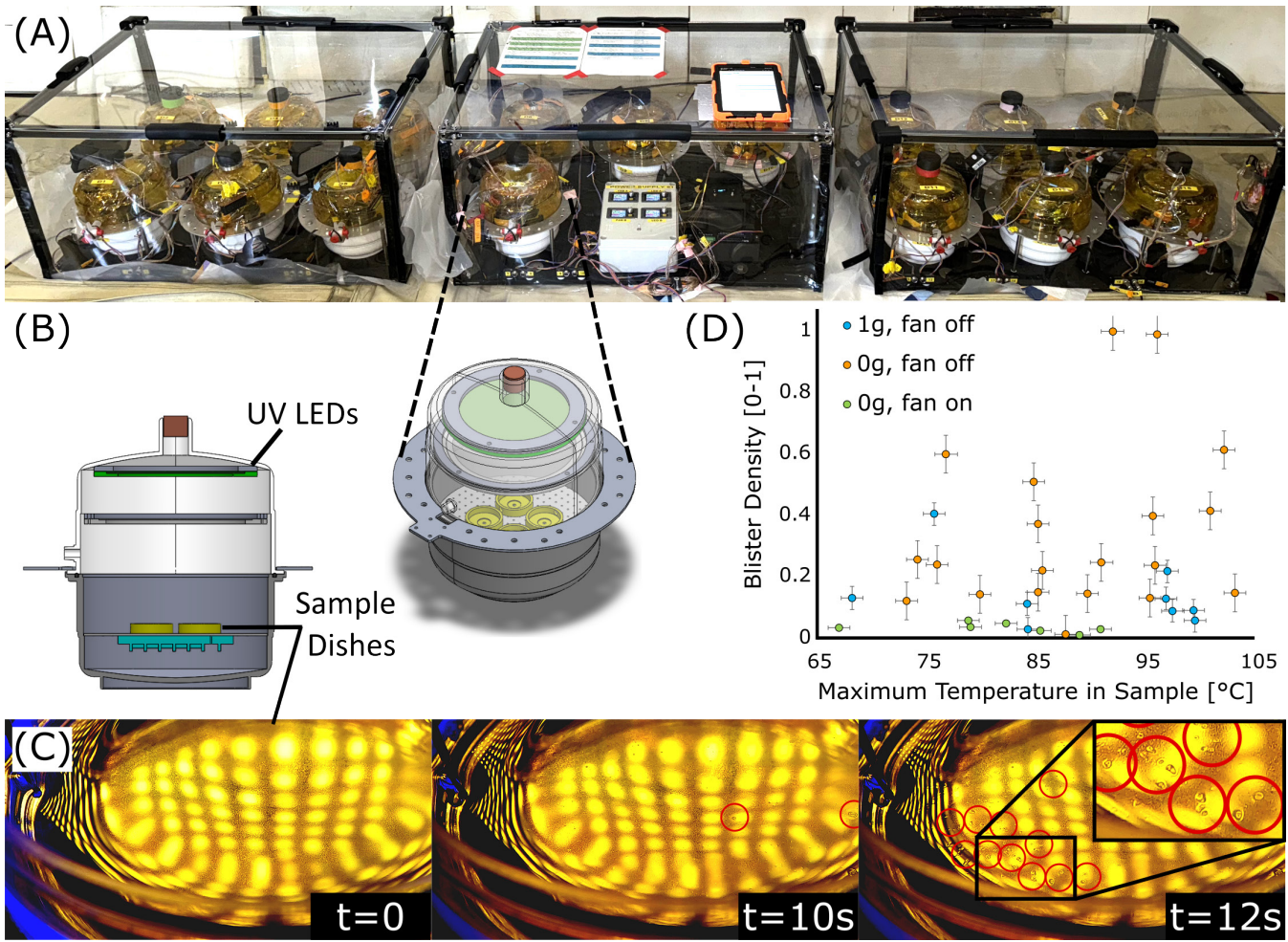

**Figure 2. Parabolic flight experiments investigating microgravity's effect on thermal buildup and defects during photopolymerization.** **(A-B)** Experimental setup onboard the Zero-G plane: We bolted 16 desiccators, each holding 2 to 4 sample dishes filled with photopolymer resins, to the plane floor. We cured the polymers in-flight using UV LED arrays mounted inside the desiccator lids. We regulated convective heat flow by activating fans mounted inside the chambers. With the fans turned off, we had natural convection under 1g and no convection in 0g. Activating the fans established consistent forced convection in both gravity regimes. We measured the temperature distribution inside the polymers with embedded thermocouples placed at predefined heights. GoPro cameras inside the desiccators recorded the evolution of thermal defects during polymerization. **(C)** Image sequence of the photopolymer surface during polymerization in microgravity: At $t = 0$, under microgravity conditions, we activated the UV LED array above the liquid polymer resin, initiating photopolymerization. At this point, the liquid surface was smooth and defect-free. At $t = 10$ s, we measured an internal temperature of 89 °C and began to spot blisters on the solidifying surface (marked with red circles). At $t = 12$ s, the temperature peaked at 92 °C. The video footage presents rapid formation of multiple overlapping blisters (marked with red circles and magnified) and the release of fumes (see Supplementary Video 4). As the appearance of the fumes coincides with the peak temperature, we suspect substantial heat loss through evaporation. **(D)** Normalized surface blister density as a function of TJ-3704A samples peak temperature and convective heat flow: blister density increases with peak temperature and is significantly more pronounced in 0g without convection. The three datasets presented correspond to natural convection during 'straight and level' flight (1g, fan off), natural convection during microgravity (0g, fan off), and forced convection during microgravity (0g, fan on). While some defects are observed in 1g conditions, microgravity exacerbates the issue greatly. Samples polymerized in microgravity exhibited both higher average temperatures and approximately double the blister count compared to those polymerized in 1g, under otherwise identical experimental conditions. Fan-induced forced convection had a pronounced cooling effect, even at the modest airflow velocities tested (~4 m/s). Average sample temperatures dropped by approximately 20 °C, and thermal defects were virtually non-existent in the fan-cooled samples.

## 2.2. Thermal model

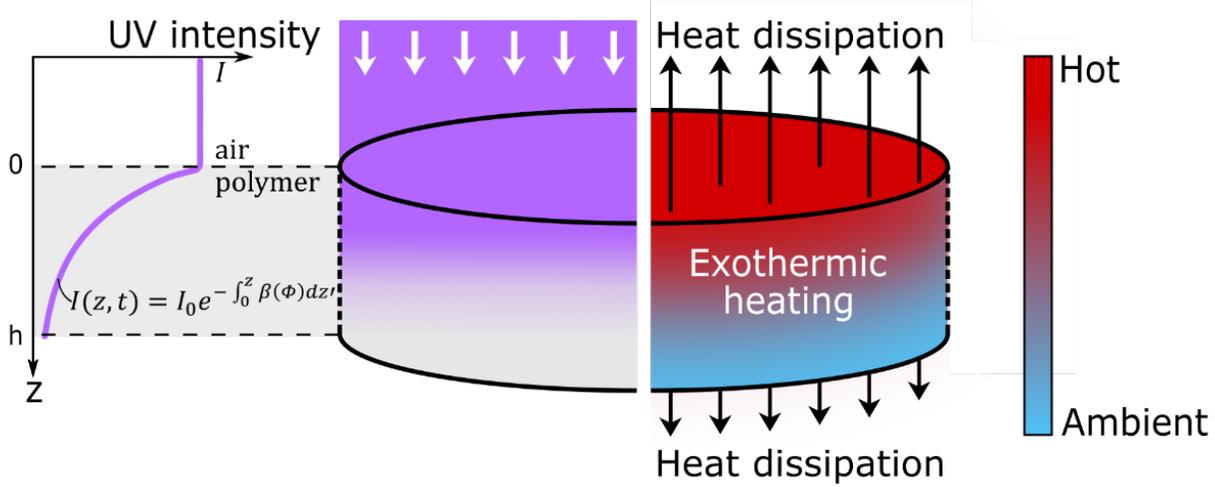

**Figure 3. Schematic illustration of the modeled system.** We consider a liquid photopolymer resin with thickness $h$, density $\rho$, heat capacity $c_p$ and thermal conductivity $k$. The photopolymer is exposed from above to UV light (white arrows) with intensity $I(x,y,z=0,t) = I_0$, which causes the polymer to undergo exothermic photopolymerization and gradually transition to a solid. The polymer's material state is modeled by a continuous state function $\Phi$ (see eq 1), such that $\Phi(z, t = 0) = 0$ for the initial liquid and $\Phi(z, t \to \infty) = 1$ for the complete solid. **Left)** As the UV light propagates through the polymer, it is gradually absorbed. We model the resulting light distribution $I(z,t)$ (purple gradient) using the Beer-Lambert law with a non-uniform absorption coefficient $\beta = \beta(\Phi)$, whose value is a function of the polymer's state. **Right)** The UV induced polymerization reactions release substantial heat, resulting in temperature gradients throughout the sample. Heat is dissipated to the environment via the top and bottom faces. Different environmental conditions are manifested through the specific thermal boundary conditions at these faces. For example, for simulations of in-lab tests, we prescribe natural heat convection at the top surface, whereas for microgravity conditions we prescribe heat conduction through the surrounding air.

We consider a liquid photopolymer resin with thickness $h$, as illustrated in Figure 3. The polymer has a density $\rho$, specific heat capacity $c_p$, thermal conductivity $k$, and UV absorption coefficient $\beta$. The liquid polymer resin is irradiated from above by UV light with uniform intensity $I(x,y,z=0,t) = I_0$. The UV light penetrates the photopolymer sample and activates the photoinitiator molecules, which initiate polymerization reactions, gradually curing the entire sample until it is completely solidified. For the majority of photopolymers, heat is released during polymerization both due to the exothermic polymerization reaction and due to the release of latent heat during phase transition from liquid to solid. Importantly, during this phase transition, material properties, such as $c_p$ and $\beta$, are also expected to evolve. For our model, we consider an initially homogenous polymer and assume that the area of interest, at the center of the polymer sample, is much smaller than the dimensions of the entire sample. Thus, we neglect any in-plane variations and develop a one dimensional model that accounts for changes in the z-direction.

To describe the progression of the polymerization reaction in space ($z$) and time ($t$), we define a continuous function $\Phi(z,t)$ as

$$\Phi(z,t) = \begin{cases} 0 & \text{Liquid resin} \\ 0 < \Phi(z,t) < 1 & \text{Partial polymerization} \\ 1 & \text{Solid polymer} \end{cases} \quad (1)$$

The rate of photopolymerization, $\frac{\partial \Phi}{\partial t}(z,t)$, is dictated by the local UV intensity distribution, $I(z,t)$. At the polymer's surface, $z = 0$, the radiation intensity is $I_0$, and it diminishes due to absorption as the light propagates through the photopolymer. Neglecting other weaker effects such as light scattering, we model the light intensity distribution using the Beer-Lambert Law,

$$I(z,t) = I_0 \exp\left(-\int_0^z \beta(\Phi) dz'\right), \quad (2)$$

where $\beta$ is the absorption coefficient. Typically, $\beta$ decreases as photoinitiators are depleted,[10] and so light distribution is expected to vary in time and space through its dependence on the polymerization state $\Phi(z,t)$.

While our model is applicable to any photopolymer, in this work, we limited our material characterization efforts to the three polymer resins that were used in our ISS experiment: TJ-3704A, NOA61 and NOA63. We characterize the absorption coefficient $\beta(\Phi)$ by measuring the transmitted intensity through liquid and solid polymer samples of varying thicknesses. In our model, we extend the measured absorption coefficients to all intermediate material states via linear interpolation, such that

$$\beta(\Phi) = \beta(0) + \big(\beta(1) - \beta(0)\big)\Phi. \quad (3)$$

In this work, we seek to model the evolution of the polymer's temperature distribution during the photopolymerization process, in various environmental conditions, including microgravity. Following our model's assumptions, and accounting for the heat released by the polymer, the energy balance on a differential unit volume yields the following one-dimensional heat equation,

$$\rho c_p \frac{\partial T}{\partial t} = \frac{\partial}{\partial z}\left(k \frac{\partial T}{\partial z}\right) + \dot{q}_{\text{chemical}} + \dot{q}_{\text{latent}} + \dot{q}_{\text{absorption}}. \quad (4)$$

Here, $T(z,t)$ denotes the polymer's temperature, $\dot{q}_{\text{chemical}}$ is the heat released due to chemical reactions, $\dot{q}_{\text{latent}}$ is the heat released due to the polymer's phase change, and $\dot{q}_{\text{absorption}}$ is the absorbed radiative heat due to the UV illumination. Through UV-DSC measurements (UV-coupled Differential Scanning Calorimetry), further discussed below, we found that for the chemical formulations and UV intensities considered in this work, the absorbed radiative heat is negligible when compared to the generated chemical heat. Hence, we neglect $\dot{q}_{\text{absorption}}$ with respect to other heat sources. Additionally, we assume that the variation in the polymer's thermal conductivity is small, an assumption that holds for most thermoset polymers due to their amorphous nature and rigid crosslinked structure,[11] and is self-consistent with our experimental results. Under these assumptions, eq. 4 simplifies to

$$\frac{\partial T}{\partial t} = \alpha \frac{\partial^2 T}{\partial z^2} + \frac{\dot{q}}{\rho c_p}, \quad (5)$$

where $\alpha = \frac{k}{\rho c_p}$ is the polymer's thermal diffusivity, and $\dot{q}(z,t) = \dot{q}_{chemical}(z,t) + \dot{q}_{latent}(z,t)$.

At the boundaries of the polymer $(z = 0, h)$, we prescribe either a heat conduction condition

$$BC1: \quad k\frac{\partial T}{\partial z} = k_{env}\frac{\partial T}{\partial z}, \quad (6)$$

where $k_{env}$ is the heat conduction coefficient of the surrounding environment, or a heat convection condition, modeled using Newton's cooling law,

$$BC2: \quad \frac{\partial T}{\partial z} = \hbar(T - T_{amb}), \quad (7)$$

where $\hbar$ is the heat convection coefficient and $T_{amb}$ is the ambient temperature. For the case of polymerization on Earth, we apply $BC1$ at $z = h$ and $BC2$ at $z = 0$, using a value for $\hbar$ that corresponds to natural convection. For the case of polymerization in microgravity, we apply $BC1$ at both boundaries, indicating no convection. When forced convection is applied (either on Earth or in space), $BC2$ is applied with a value for $\hbar$ that is defined by the flow velocity.

When considering the thermal governing equation (eq. 5), it is important to recall that the material properties, $\alpha$ and $c_p$, as well as the generated heat, $\dot{q}$, are all expected to evolve with the material state, $\Phi(z,t)$, and temperature, $T(z,t)$. We characterized the polymers' thermal diffusivity, $\alpha$, by applying the laser flash analysis (LFA) technique (implemented using a NETZSCH LFA-457 *MicroFlash* apparatus) and found negligible variation in the diffusivity at different temperatures and polymerization states (see measurement details in the SI). We thus model the thermal diffusivity as a material specific constant

$$\alpha(\Phi, T) = \alpha_0. \tag{8}$$

We measured the polymers' specific heat capacities, $c_p(\Phi, T)$ by DSC (Mettler Toledo DSC 3) in both liquid ($\Phi = 0$) and solid ($\Phi = 1$) phase and for temperatures ranging from 25 to 100 °C, by heating at a rate of 10 °C/min under a nitrogen atmosphere. For all the polymers measured, we found that in both liquid and solid state, $c_p$ values rise linearly with temperature. At a given temperature, $c_p$ values of the solids were approximately 20% lower than those of the corresponding liquids. To interpolate the $c_p$ values of intermediate phase states, we assume a linear dependence such that

$$c_p(\Phi, T) = c_p(\Phi = 0, T) + \left(c_p(\Phi = 1, T) - c_p(\Phi = 0, T)\right)\Phi, \tag{9}$$

where the functions $c_p(0, T)$ and $c_p(1, T)$ are defined by linear fits of the measured data. Detailed measurements are provided in the SI.

To model the rate of polymer heat release, $\dot{q}(z,t)$, we must account for the underlying polymerization kinetics. Ideally, one would develop a mechanistic cure-kinetic model, such as that presented by O'Brien *et. al.*[12]. Such models describe the combined effects of chemical properties, such as reactant concentrations, molecular diffusion coefficients and reaction rates, on the polymerization progression. As such, these models are very useful for obtaining insight into the polymerization process and informing the design of custom chemical formulations. However, for the same reason, the application of mechanistic models to predict the behavior of commercial polymer resins is limited as they require detailed and proprietary chemical information which is difficult to obtain. Instead, we adopt phenomenological modeling of the heat released during polymerization following the approach of Bartolo *et. al,*[13] where the total generated heat is related to the rate of polymerization,

$$\dot{q}(z,t) = \rho \Delta H \frac{\partial \Phi}{\partial t}(z,t). \tag{10}$$

Here $\Delta H$ is the total energy per unit mass released during full polymerization. We characterized the heat generation for each of our polymers using a UV-DSC system (Mettler Toledo DSC-1 coupled with Hamamatsu LC8-02 UV light source and a photocalorimetry kit), i.e., we placed ~10 mg photopolymer resin samples inside dedicated 40 µL aluminum crucibles and illuminated them from above using various intensities of UV light at various isothermal temperatures, in a nitrogen atmosphere, while measuring the heat release as a function of time. Figure 4A presents such measurements of TJ-3704A after a standard baseline subtraction. Integrating these measurements in time yields the total released heat per unit mass, $\Delta H$. Figure 4B presents the same data as a function of the polymerization state $\Phi$, based on the integration of Figure 4A, and shows typical autocatalytic behavior, where the initially slow reaction accelerates and then decelerates as reactant mobility becomes hindered by the increasing polymer viscosity. The measurements indicate that illumination intensity significantly affects the rates of reaction and extent of released heat. In contrast, within our range of interest, temperature has a much smaller influence on the rate of reaction. For these reasons, we adopt the autocatalytic model introduced by Kamal et al.[14–16] in combination with the power law dependence on intensity proposed by Maffezzoli *et al.*[17], while neglecting temperature dependence:

$$\dot{q} = \rho \Delta H \frac{\partial \Phi}{\partial t} = \dot{q}_0 \left(\frac{I}{I_{ref}}\right)^b \Phi^m (1 - \Phi)^n, \tag{11}$$

where $b$, $m$, $n$ and $\dot{q}_0$ are coefficients to be determined by data fitting, and $I_{ref}$ is an arbitrary reference value for the intensity.

Figure 4C presents the total released heat per unit mass, $\Delta H$, as a function of the UV intensity, $I$, for all three characterized polymers, as measured at 25 ˚C isothermal conditions. For a sufficiently high intensity (above the marked $5I_{sat}$) the total heat generation plateaus at a value we denote as $H_\infty$, which has been previously shown[18] to correlate with a high degree of polymer conversion. However, consistent with several previously reported works[17,19,20], below this threshold we measure a strong dependence of the total released heat on intensity, which is indicative of a decrease in the attained degree of conversion. This range is particularly relevant for our case, where light is significantly attenuated in deeper layers of the polymer. As shown in the plot, this behavior is well described by a sigmoid curve,

$$\Delta H(I) = \frac{H_\infty}{1 + \left(\frac{H_\infty - H_0}{H_0}\right) e^{-\frac{I}{I_{sat}}}}, \tag{12}$$

where $H_0, H_\infty$ and $I_{sat}$ are fitting parameters.

The degree of conversion itself is of particular interest as it defines many macroscopic properties such as tensile strength and refractive index.[21] While $\Phi$ describes the progression of the polymerization reaction, it does not accurately predict the degree of conversion of weakly illuminated polymers. Following Rusu *et al's.*[18] convention, we define the degree of conversion $\theta(z, t)$ as,

$$\theta(z, t) = \frac{1}{\rho H_\infty} \int_0^t \dot{q}(z, t') dt' = \frac{\Delta H(I)}{H_\infty} \Phi(z, t). \tag{13}$$

As we show in the Results, this implies that a polymer can be fully solidified ($\Phi(z, t \to \infty) \to 1$) while still exhibiting a spatially varying final degree of conversion, $\theta(z, \infty)$. This spatial dependence, which arises from nonuniform UV exposure during photo-curing, leads to different regions of the material irreversibly developing different macroscopic properties.

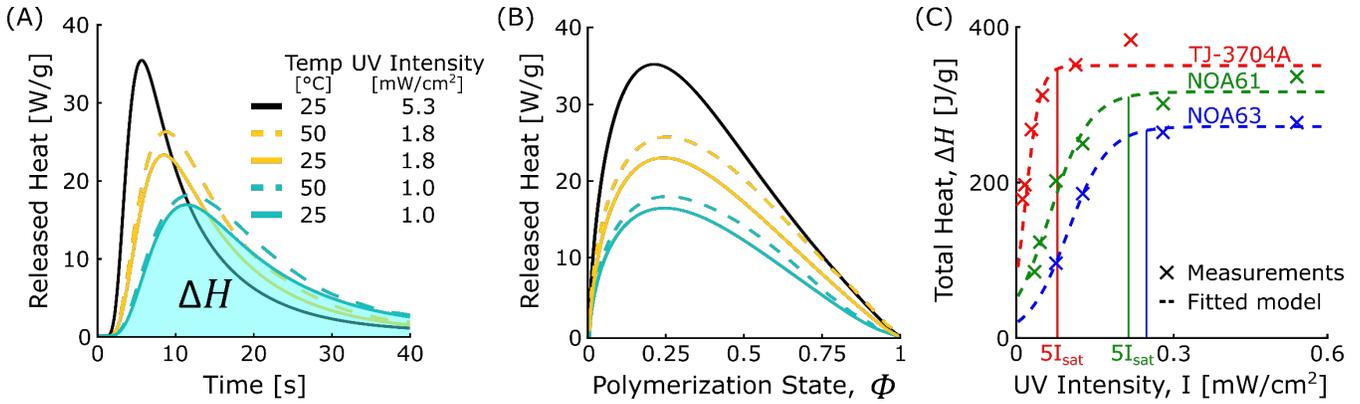

**Figure 4. Modelling of photopolymer heat release based on UV-DSC characterization. (A)** Measurements of TJ-3704A heat release (per unit mass) during photopolymerization under various UV intensities and isothermal temperatures, after a standard baseline removal. The area under each curve indicates the total energy (per unit mass), *ΔH*, released during the reaction (illustrated by the light blue region for the case of 25 °C and 1.0 mW/cm$^2$). The results show a strong dependance of the instantaneous heat-release, $\dot{q}$, on the intensity of the incident UV radiation with a weaker dependance on isothermal temperature. **(B)** Plotting the same heat release measurements against the polymer reaction state *Φ*, based on integration of the curves in A, reveals a typical non-monotonic autocatalytic behavior, where the reaction rate initially accelerates, reaches a peak, and then decelerates. As shown in the Supplementary Information, this behavior is well captured by our autocatalytic model (eq. 11). **(C)** Measurements of the total released heat (per unit mass) of all three polymers as a function of UV intensity. Beyond a certain light intensity (above the marked $5I_{sat}$) the total heat generation saturates to its maximum value, whereas below this threshold the released heat is highly dependent on light intensity. This behavior is well described by a sigmoidal curve (eq. 12), as shown in the plot.

## 3. Results and Discussion

### 3.1. Model results and experimental validation

In designing a photopolymerization protocol, one typically searches for the combination of governing system parameters such as UV exposure intensity and duration, polymer material properties and thickness, and thermal boundary conditions, that allow for fast polymerization while ensuring that the resulting elements are both homogeneously polymerized (uniform degree of conversion) and defect-free, requiring temperature distributions to remain below critical thresholds throughout the entire process.

As described in Section 2.2, our model predicts the temporal and spatial evolutions of the generated heat and polymerization rate (eq. 10-13), accounting for UV light distribution (eq. 2), the polymer's material properties (eq. 3, 8, 9), and various thermal boundary conditions (eq. 6, 7). To simulate a fabrication process and obtain the time and space dependent temperature and polymerization state maps, we numerically solve this system of equations. We first discretize the spatial differential operators using second-order central finite differences. Following a grid dependency test, we used N = 1000 nodes, which corresponds to 5 μm resolution for a 5 mm sample thickness. This results in a system of first-order differential equations in time, which we integrate using MATLAB's ODE15s solver. For simulations that prescribe heat conduction boundary conditions (BC1, eq. 6), we extended the computational domain to include an external 50 mm thick region representing the surroundings with a conductivity $k_{env}$, and a fixed temperature boundary condition of 23 °C applied at its outer edge.

Figure 5 presents simulation results for the case of a 6 mm TJ-3704A sample (similar to the polymer thicknesses used for our lens fabrication) undergoing photopolymerization driven by exposure to 365 nm UV light from above, with an intensity of 2.9 mW/cm$^2$ for 1000 seconds. In this case, we simulate polymerization on Earth, and thus the base of the sample, $z = 6$ mm, is thermally insulated with Styrofoam while its top, $z = 0$, is cooled by natural convection. Surprisingly, under these typical conditions, the resulting temperature map (Figure 5A) predicts the development of temperatures as high as 110 °C. The upper polymer layers, which receive the strongest UV radiation, reach their maximum temperatures within the first minute of exposure, while at later times (t = 200 s) a substantial thermal buildup also occurs farther away from the polymer's surface due to the polymer's low thermal conductivity and thermal insulation of the chamber base.

Figure 5B depicts the evolution in time of the polymerization reaction state $\Phi$, with 0 denoting the initial liquid polymer resin and 1 denoting a fully solidified polymer. Due to the strong UV absorption, the polymer farthest away from the surface is predicted to be only partially polymerized, even after the relatively long 1000 s of continuous exposure.

Figure 5C depicts the evolution in time of the polymer's degree of conversion $\theta$. The plot demonstrates that due to diminishing light intensity at deeper layers, and regardless of the exposure time allowed, the solid regions of the polymer (i.e. where $\Phi = 1$ in Figure 5B) reach very different degrees of conversion. This is expected to result in nonuniform macroscopic properties of the solid, such as mechanical strength and index of refraction. Such insight is important for designing better polymerization conditions, avoiding common unbased assumptions that sufficient illumination time at a lower intensity would ultimately yield to uniform conversion, or that the total dose is a governing parameter for the resulting material properties. Using the model, one can address the inherent tradeoff: illuminating at an intensity much higher than $5I_{sat}$ would theoretically yield a uniform material, but the temperature developed in the process would be excessive. On the other hand, illuminating at a low intensity can avoid extreme temperatures, but necessarily result in regions of different polymer conversion.

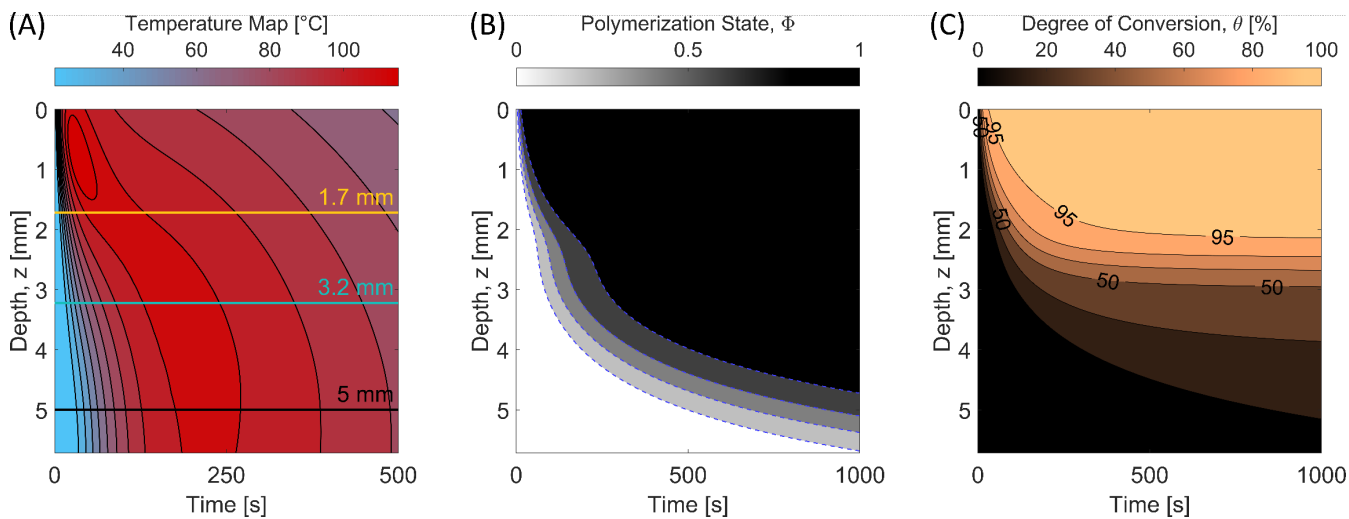

**Figure 5. Simulated curing of a 6 mm thick TJ-3704A photopolymer sample exposed to UV from above. (A)** Temperature distribution during polymerization under natural convection cooling from the free surface, $z = 0$, and thermal insulation at the bottom surface, $z = 6$ mm. The simulation reveals a highly nonuniform temperature map which is a result of the polymer's heat generation being strongly dependent on light intensity (which undergoes strong absorption), as well as of its low thermal conductivity. High temperatures, well above 100 °C, are predicted near the polymer's surface (~1 mm depth) where the relatively strong UV intensity triggers very rapid polymerization and heat release. While the UV intensity in the deeper layers is significantly lower, at later times (~200 s) the bulk of the sample (~5 mm from the surface) also exhibits substantial thermal buildup due to the polymer's poor thermal conductivity and the insulation boundary condition. The horizontal lines correspond to the depths at which the experimental measurements presented in Figure 6 were taken. **(B)** Predicted progression of polymerization, showing the transition from liquid resin ($\Phi = 0$, white) to fully solid polymer ($\Phi = 1$, black). The model successfully captures the gradual polymerization process, where material close to the free surface has already solidified yet the polymer bulk remains liquid. **(C)** Predicted degree of polymer conversion. The model predicts that while the TJ-3704A sample

may be completely solidified after ~1000 s of curing, for the simulated UV intensity and polymer thickness, the resulting degree of conversion will be highly non-uniform. As a result, the distribution of macroscopic properties such as refractive index are expected to be accordingly nonuniform. The simulation was performed for a light intensity of 2.9 mW/cm², at a wavelength of 365 nm.

We experimentally validated our model by comparing the thermal predictions of Figure 5A to time dependent temperature measurements taken in the lab under the same conditions, as shown in Figure 6. We filled a 30 mm diameter petri dish with TJ-3704A photopolymer resin to a height of 6 mm and placed it into a matching cavity within a large block of Styrofoam for thermal insulation, leaving the top exposed. We then irradiated the polymer with 365 nm UV light from above at an intensity of 2.9 mW/cm², measured at the polymer's free surface ($I_0$) with a calibrated power meter (S120VC, Thorlabs). Throughout the experiment we measured the temperature at depths of 1.7, 3.2 and 5 mm (indicated by the horizontal lines in Figure 5A) with K type thermocouples, sampled at 25 Hz. Two additional K-type thermocouples measured the ambient temperature at two points, each approximately 1 cm away from the dish.

Figure 6B compares the experimental temperature measurements with the corresponding model predictions, obtained without any fitting parameters. Overall, the model demonstrates strong agreement with the experiments, successfully capturing the temporal evolution of temperature at multiple depths within the polymer both during the active heat generation at early times and under cool-down at long times. At shallow depths near the polymer–air interface (yellow curve, z = 1.7 mm), the measured thermal behavior differs from that observed deeper in the material. While the model accurately predicts the initial heat-up and the long-time cooling of this near-surface region, it overestimates the peak temperature during the intermediate times when the surface reaches its maximum temperature. This discrepancy is localized to the regions closest to the free surface and does not appear at greater depths. Given that the TJ-3704A resin formulation contains between 5 to 15 wt.% of HPMA (hydroxypropyl methacrylate), a component known to be volatile at the measured temperature range[22], we hypothesize that evaporative cooling, neglected in our model, is responsible for the reduced measured temperature at the interface. This explanation is consistent with the substantial surface blistering observed on this sample and falls in line with our parabolic-flight observations where excess temperatures correlated with the formation of blistering (Figure 2).

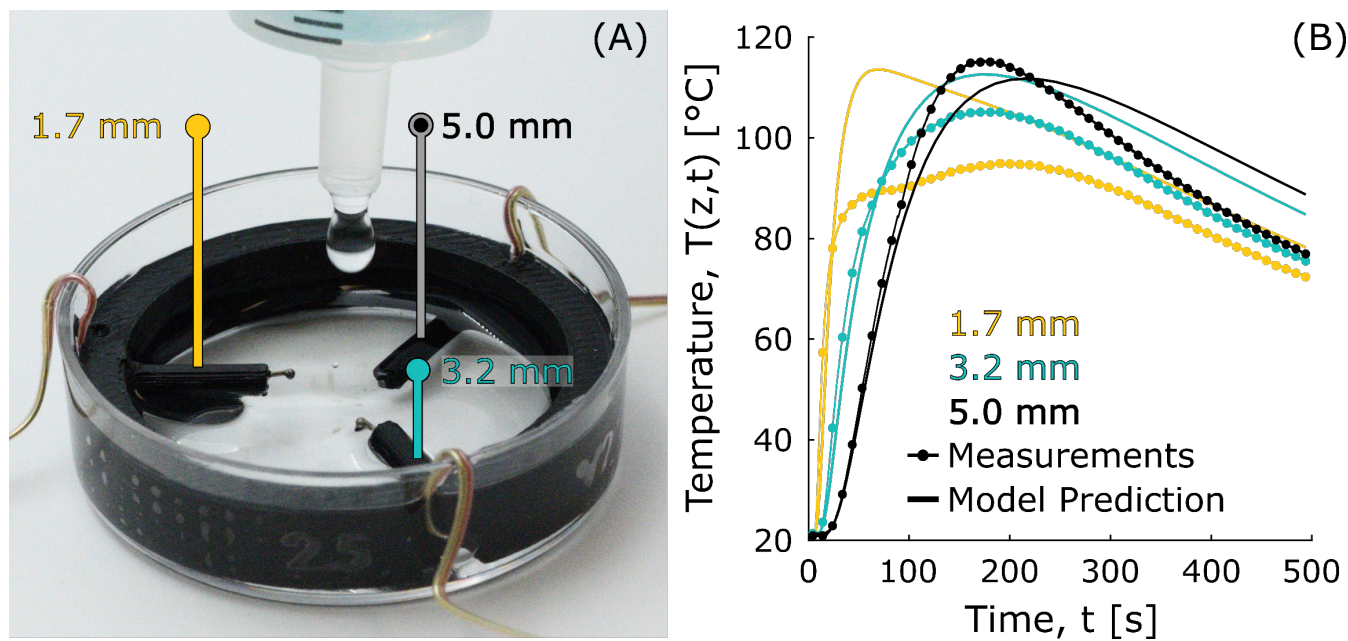

**Figure 6. Experimental validation of the thermal model. (A)** To measure the temperature evolution during polymerization, we embedded three K-type thermocouples within a Petri dish. When the dish was filled with polymer resin to a height of 6 mm, the thermocouples were located at depths of 1.7, 3.2 and 5 mm under the surface. During the experiment, the dish was placed into a matching cavity within a Styrofoam block for insulation, with the top surface remaining exposed to the atmosphere and UV radiation. **(B)** Comparison of experimental temperature measurements (circles) and model predictions (thick solid lines) at the three measured depths, during the polymerization of a TJ-3704A sample. The polymer was cured with 365 nm UV light with an intensity of 2.9 mW/cm², measured at the free surface. The thermocouple measurements, sampled simultaneously at 25 Hz, are presented by the thin lines. For visual clarity, the presented data points (circles overlayed on the thin lines) are a subset of the full dataset, spaced 10 s apart. The presented model predictions (thick solid lines) correspond to the horizontal lines marked in Figure 5. The model shows good agreement with the experiments, predicting well the rate of temperature rise at early times ($t < 50$ s), the maximum temperature reached, and the cool-down slope at long times ($t > 250$ s). At intermediate times, where the surface temperature is at its peak, the measured temperature profile close to the surface ($z = 1.7$ mm, yellow) deviates from our model predictions and exhibits behavior different from that measured at the two larger depths. Given that the TJ-3704A resin is comprised of 5 to 15% HPMA (Hydroxypropyl methacrylate), which is volatile at these temperatures[22], we hypothesize that this discrepancy is the result of evaporation, which our model neglects.

### 3.2. From Observation to Prediction: Modeling the ISS Experiment

To demonstrate the capabilities of the predictive model, we show its use in explaining the blistering observed in our ISS experiments, and in designing better experimental conditions that would avoid such defects.

Thermal blistering is a result of mass loss through evaporation and boiling at the liquid-air interface,[8] (see Supplementary Video 3). As shown in Figure 7A, to characterize the critical temperatures above which such mass loss can be expected, we performed a thermogravimetric analysis (TGA) of the three resins previously used in the ISS experiment. We found that each polymer has a unique temperature above which significant mass loss is observed. Both NOA61 and NOA63 Norland adhesives lose mass from about 175 °C whereas TJ-3704A already loses 5% of its mass at a relatively low temperature of 120 °C.

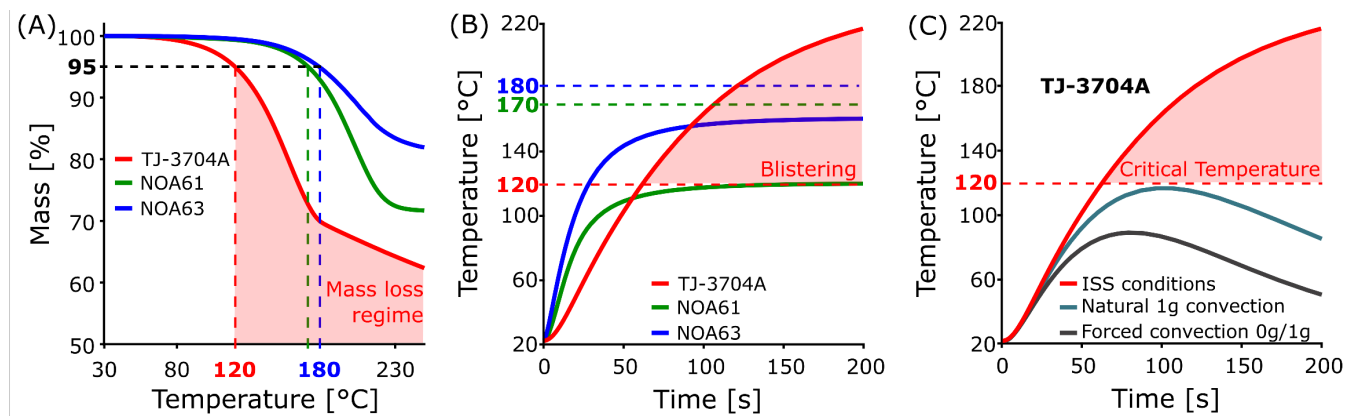

**Figure 7. Revisiting our ISS experiments as a case study for our thermal model. (A)** Thermogravimetric measurements of TJ-3704A, NOA61, and NOA63, showing mass loss as a function of temperature. The dashed lines indicate the critical temperatures at which the polymers' mass dropped below 95% of its initial value. **(B)** Model results showing the evolution of temperature with time for the three polymers, under the conditions of the ISS experiment: microgravity environment, 4 mm sample thickness, simultaneous UV exposure from both sides with an intensity of 0.455 mW/cm². Only TJ-3704A exceeds its critical temperature, consistent with the blistering observed in the ISS experiments. **(C)** Simulated temperature profiles for TJ-3704A under three heat transfer conditions: microgravity without convection (ISS conditions, red), 1g natural convection (teal), and fan-induced forced convection (black). Comparing 0g and 1g conditions illustrates how the lack of convection drives the system to temperatures that exceed the polymer's threshold. The simulation also demonstrates that for this particular case, introducing a fan with wind speeds as low as 5 m/s effectively prevents thermal defects. This finding is supported by our parabolic flight experiments.

The lenses produced in the ISS experiments had a characteristic thickness of 4 mm and were cured from both sides simultaneously by 365 nm UV light with an intensity of 0.455 mW/cm$^2$. During their curing, the polymerization chamber was closed, inhibiting any airflow within the chamber. This, together with the absence of gravity, implies that no heat could be lost from the lens through convection. We thus modelled heat dissipation out of the lens through pure conduction[23], with a reference temperature of 23 °C at a distance of 50 mm from the surface of the lens.

As shown in Figure 6B, our simulations indicate that the TJ-3704A lenses exceeded their critical temperature threshold (as found by the TGA) approximately one minute into curing, consistent with video footage of the experiment showing the escape of fumes from the curing chamber when it was opened after 4 min of curing (see Supplementary Video 2), and with the severe surface blistering observed on the returned ISS samples (Figure 1C). The model also shows that the NOA resins remained below their respective thresholds, which again matches the experimental outcomes in which these lenses were defect-free.

The use of the model is, of course, not limited to post-experimental investigations and can be used as a design tool for future in-space manufacturing processes. With the model, we can simulate variations of experimental conditions such as polymer thickness and light intensity variation over time. We can also investigate desired material properties and their role in thermal regulation as well as thermal control through external conditions such as forced convection or reduced environmental temperatures. As an example, we used the model to quantify what magnitude of forced convection would be needed to avoid the onset of thermal defects under otherwise identical experimental conditions. The black curve in Figure 6C presents the case where the polymer-air interface is subjected to a Newton cooling law boundary condition with a forced heat transfer coefficient of 50 W/(m$^2$K). This translates to a laminar airflow of approximately 5 m/s, which can be easily achieved with a small fan. The simulation indicates that such forced convection would be sufficient to prevent the overheating of the TJ-3704A samples. This result is supported by our experimental findings from the parabolic flight tests.

## 4. Conclusions and Outlook

Future long-duration space missions, such as crewed Mars expeditions, will require in-situ, on-demand manufacturing capabilities. Photopolymer-based manufacturing processes are promising as they inherently require little energy and simple infrastructure, and as liquid-resin cargos are very volume-efficient. However, heat released during rapid photopolymerization can lead to mechanical defects if left unmanaged. Regulating this heat is generally challenging due to the polymer's low thermal conductivity and particularly challenging in microgravity, where natural convection does not occur.

Photopolymer resins are complex mixtures of monomers, photoinitiators, diluents, inhibitors, and other additives. As a result, predicting their thermal behavior during polymerization is challenging, and usually requires detailed knowledge of proprietary formulations. In this work, we demonstrated an experimental characterization method that bypasses the need for such prior knowledge. Specifically, we propose measuring both the rate of heat generation and the total heat released through UV-DSC measurements. These measurements reveal that material response is highly sensitive to UV intensity, implying that accurate process modeling must account for spatial variations in light distribution within the sample. This challenge is further compounded by the strong absorption of UV light in such polymers, which results in intensity gradients in the material.

We developed a predictive thermal model and validated it using laboratory and parabolic flight data. By integrating heat transfer, light absorption, and evolving material properties, the model predicts temperature evolution during curing under various environmental conditions, including microgravity. Simulations based on the model can serve as a practical tool for defect-free in-space manufacturing, supporting the transition from Earth-based to in-space methods. For instance, when applied to our ISS experiments, it accurately explained TJ-3704A resin failures and demonstrated the effectiveness of forced convection as a strategy to prevent overheating in this particular process.

The model does not include heat loss through evaporation or boiling. While including such effects could extend the validity of the model, they are not strictly necessary as the desired work point of practical applications is one at which temperatures remain below their critical threshold and defects associated with these phase changes are avoided. If additional accuracy is desired at borderline temperatures, such effects should be included.

The solutions we presented are one-dimensional and assume uniform illumination with material absorption that follows the Beer-Lambert law. Such modeling is directly applicable to photopolymerization fabrication methods such as polymer casting, coating, or our Fluidic Shaping which utilize UV exposure over large areas to flood-cure elements. The model can be adapted for other methods such as DLP and SLA printing by considering a non-uniform intensity map at the boundary, $I(x, y, z = 0, t)$ and by generalizing the heat equation (eq. 5) to include in-plane heat conduction. Methods such as VAM, which modulate light intensity through interference, would have to include this intensity modulation in the source term of the thermal model. Despite being one-dimensional, the model is useful for designing the allowable working conditions for avoiding defects or boiling in three-dimensional objects being polymerized. The case we have focused on is that of lenses, which have a very regular axisymmetric shape. However, we expect the model to be equally relevant for more general shapes, as long as their dimensions are significantly larger than the absorption depth. For such objects, the polymerization can be considered to occur on a thin outer 'shell' of the object, and thus still be represented well by the 1D model. However, if one desired an accurate prediction of the temperature distribution in time and space, both the heat and UV propagation models should be expanded to 3D. In such a 3D heat model, the heat conduction operator will naturally take its operator form, but the source term in eq. 5 will remain unchanged.

In summary, our findings emphasize that effective thermal management is essential for reliable in-space photopolymer manufacturing. The validated model provides a practical design tool needed to adapt Earth-based photopolymerization processes for application in space, and even more so for the development of microgravity-based methods that cannot be tested on Earth. We believe that such models and simulations are essential for paving the way toward scalable and dependable in-space manufacturing for future long-duration missions.

## Data Availability

The data is available upon request.

## Code Availability

The numerical solver is available upon request.

## Funding

This project has received funding from the European Research Council under the European Union's Horizon 2020 Research and Innovation Programme, grant agreement 10104451 (Fluidic Shaping). Views and opinions expressed are, however, those of the author(s) only and do not necessarily reflect those of the European Union or the European Research Council Executive Agency. Neither the European Union nor the granting authority can be held responsible for them. We also acknowledge financial support from the Israeli Space Agency within the Israeli Ministry of Innovation, Science and Technology, and from the Norman and Helen Asher Space Research Institute (ASRI) fund. We also thank NASA Ames Center Innovation Fund for its financial support of this project and the ISS National Lab for providing material safety certification at the NASA White Sands Test Facility. We are grateful to NASA Flight Opportunities program and NASA Innovative Advanced Concepts program (both within NASA's Space

Technology Mission Directorate) for funding the parabolic flight experiments. M.E. was supported by the Ramon Graduate Fellowship of the Israel Ministry of Innovation, Science and Technology. I.G. acknowledges the support of ISEF and is grateful to the Azrieli Foundation for the award of an Azrieli Fellowship.

## Acknowledgments

We thank RAKIA for creating the opportunity for experiments on the ISS. We thank Aliza Shultzer for managing the logistics of the parabolic flight experiments. We also thank Howard Cannon for coordinating flight certification of the experimental equipment for parabolic flights, managing flight preparation logistics, participating in the assembly and installation of the experimental equipment aboard the Zero-G aircraft, and performing the in-flight experiments.

## Author Contributions

E. Stibbe performed the ISS experiments. D.W. conceived the working hypothesis and performed preliminary experiments. J.E., D.W., M.E., Y.K., O.L., K.H., A.R., I.G. and V.F developed the parabolic flight experimental hardware. J.E., D.W., M.E., Y.K., O.L., K.H., A.R., I.G., T.S., R.T., and E.B. performed the parabolic flight experiments. J.E., H.A.H., E.S., M.E, and Y.A. performed the polymer material characterization experiments. J.E. and M.B. developed the model. J.E. wrote the numerical solver and performed validation experiments. J.E., M.E., Y.K., and M.B. analyzed the data. J.E., A.H., and M.B. wrote the manuscript while all the authors reviewed and commented on it.

## Competing Interests

The authors declare no competing financial or non-financial interests.

## References

1. Prater, T. *et al.* 3D Printing in Zero G Technology Demonstration Mission: complete experimental results and summary of related material modeling efforts. *Int. J. Adv. Manuf. Technol.* **101**, 391–417 (2019).

2. Kringer, M., Böhrer, C., Frey, M., Pimpi, J. & Pietras, M. Direct Robotic Extrusion of Photopolymers (DREPP): Influence of microgravity on an in-space manufacturing method. *Front. Space Technol.* **3**, (2022).

3. Waddell, T. *et al.* Use of volumetric additive manufacturing as an in-space manufacturing technology. *Acta Astronaut.* **211**, 474–482 (2023).

4. Rangroo, A. Auxilium Biotechnologies 3D prints medical devices in Space. *3D Printing Industry* https://3dprintingindustry.com/news/auxilium-biotechnologies-3d-prints-medical-devices-in-space-236152/ (2025).

5. 3D Printing in Space - Photocentric. https://photocentricgroup.com/3d-printing-in-space/.

6. Frumkin, V. & Bercovici, M. Fluidic shaping of optical components. *Flow* **1**, E2 (2021).

7. Luria, O. *et al.* Fluidic shaping and in-situ measurement of liquid lenses in microgravity. *Npj Microgravity* **9**, 74 (2023).

8. Albalak, R. J., Tadmor, Z. & Talmon, Y. Polymer melt devolatilization mechanisms. *AIChE J.* **36**, 1313–1320 (1990).

9. Luria, O. *et al.* In-space manufacturing of optical lenses: Fluidic Shaping aboard the International Space Station. Preprint at https://doi.org/10.48550/arXiv.2510.06474 (2025).

10. Stiles, A., Tison, T.-A., Pruitt, L. & Vaidya, U. Photoinitiator Selection and Concentration in Photopolymer Formulations towards Large-Format Additive Manufacturing. *Polymers* **14**, 2708 (2022).Page 17 of 18

# Supplementary Information

# Modeling the Thermal Behavior of Photopolymers for In-Space Fabrication

Jonathan Ericson[a], Daniel Widerker[a], Eytan Stibbe[b], Mor Elgarisi[a], Yotam Katzman[a], Omer Luria[a], Khaled Gommed[a], Alexey Razin[a], Amos A. Hari[a], Israel Gabay[a,‡], Valeri Frumkin[a,§], Hanan Abu Hamad[c], Ester Segal[c], Yaron Amouyal[d], Titus Szobody[e], Rachel Ticknor[f], Edward Balaban[f] & Moran Bercovici[a,g,*]

[a] Faculty of Mechanical Engineering, Technion - Israel Institute of Technology, Haifa, Israel
[b] The Rakia Mission, 3 Shadal St. Tel Aviv-Yafo, Israel
[c] Faculty of Biotechnology and Food Engineering, Technion - Israel Institute of Technology, Haifa, Israel
[d] Faculty of Materials Science and Engineering, Technion - Israel Institute of Technology, Haifa, Israel
[e] Department of Chemical and Biomolecular Engineering, Rice University, Houston, TX
[f] NASA Ames Research Center, Moffett Blvd., Moffett Field, CA, USA
[g] Department of Materials, ETH Zürich, Switzerland
[‡] Current affiliation: School of Chemical and Biomolecular Engineering and School of Integrative Plant Science, Cornell University, Ithaca, NY, USA
[§] Current affiliation: Department of Mechanical Engineering, Boston University, Boston, MA, USA

[*] Corresponding author: mberco@technion.ac.il

**S1. Supplementary Videos (YouTube links)**

Video1. Summary of ISS Fluidic shaping experiment

Video2. Observations of photopolymer blistering and fume release in microgravity

Video3. Recreation of polymer blistering in the lab

Video4. Parabolic flight photopolymerization experiment

We here report the material characterization measurements of: TJ-3704A, a commercial acrylate-based resin also sold under the brand name 'VidaRosa' (Dongguan Tianxingjian Electronic Technology Co., Ltd, Guangdong, China), and NOA61 & NOA63 (Norland Optical Adhesive, Norland Products Incorporated, New Jersey, USA), which were used in this work.

**S2. Thermal diffusivity, $\alpha$, characterization by Laser Flash Analysis (LFA)**

Laser flash analysis (LFA) is a widely used transient technique for measuring a material's thermal diffusivity and is considered one of the principal methods for thermal transport characterization in solids and liquids. In this method, a short laser pulse is applied to one face of a small (dia. ranging between ca. 6 to 25 mm), planar sample, and the resulting temperature rise is recorded on the opposite face as a function of time with an infra-red detector. By fitting the rear-face temperature response to an analytical heat-diffusion solution considering diverse conditions of heat loss (conduction through the samples' faces, radial edges, radiation, etc.) and pulse-time corrections, the through-thickness thermal diffusivity, $\alpha$, can be extracted.[1]

In this work, we used LFA to characterize the thermal diffusivity of the photopolymer resins in both liquid and solid states at isothermal temperatures of 25 and 50 °C, by applying short (0.5 ms) Nd-YAG laser pulses. We implemented the measurements using a NETZSCH LFA-457 *MicroFlash* ® apparatus. We tested cylindrical polymer samples with a diameter of 12.7 mm, and a nominal thickness of 1.65 mm. Liquid polymer resin samples were measured with the use of dedicated aluminum sample containers (NETZSCH catalog number 6.256.4-91.6.00). Following typical procedure, we spray-coated all the samples front and rear faces with black carbon to ensure uniform laser absorption and emissivity and applied the Cape-Lehman method to correct for heat losses from both back and front faces of the samples as well as for the effects of finite pulse durations.[2] To assess measurement uncertainties and improve accuracy, we measured each sample N = 3 times at each temperature and material state. Our results are reported in Table S1. For the temperatures measured we found a negligible variation in the sample's thermal diffusivity. We found that the liquid polymer samples have a larger thermal diffusivity than their solid counterparts, however, as this variation is small, it leads to negligible effects on our thermal model simulation predictions. Thus, for simulations of photopolymer curing of these specific polymers, we treat the thermal diffusivity, $\alpha$, as a material specific constant,

$$\alpha(\Phi, T) = \alpha_0, \quad (S1)$$

where $\Phi$ is the material phase and $T$ is its temperature. Future applications of our model to other photopolymer formulations should begin with similar material characterization. If such measurements reveal a strong variation in thermal diffusivity, we suggest expanding the modeled thermal diffusivity to a more general form by assuming a linear dependance on the material state $\Phi$,

$$\alpha(\Phi, T) = \alpha(\Phi = 0, T) + \big(\alpha(\Phi = 1, T) - \alpha(\Phi = 0, T)\big)\Phi. \quad (S2)$$

**Table S1: Photopolymer thermal diffusivity (in mm²/s) as measured by LFA at 25 and 50 °C.** We measured the polymers in both liquid and solid phase. Each sample was measured N = 3 times at each temperature and material state with the mean and standard deviation of these measurements reported below.

**Photopolymer thermal diffusivity, $\alpha$ $mm^2/s$**

|  | TJ-3704A | | | | NOA61 | | | | NOA63 | | | |
|---|---|---|---|---|---|---|---|---|---|---|---|---|
|  | Solid | | Liquid | | Solid | | Liquid | | Solid | | Liquid | |
| Temperature °C | mean | STD | mean | STD | mean | STD | mean | STD | mean | STD | mean | STD |
| 25 | 0.086 | 0.003 | 0.145 | 0.055 | 0.097 | 0.004 | 0.163 | 0.049 | 0.078 | 0.014 | 0.224 | 0.006 |
| 50 | 0.073 | 0.006 | 0.023 | 0.084 | 0.084 | 0.004 | 0.14 | 0.022 | 0.073 | 0.013 | 0.202 | 0.009 |

## S3. Specific heat capacity, $c_p$, characterization by Differential Scanning Calorimetry (DSC)

Differential Scanning Calorimetry (DSC) is a well-established technique for determining the specific heat capacity, $c_p$, of polymers and other materials as a function of temperature. In DSC, we simultaneously heat a sample and a reference material with a well-known heat capacity, while monitoring the heat flow in time into each one.[3]

In this study, we preformed DSC measurements (with a Mettler Toledo DSC 3) of our three photopolymer formulations in both liquid and solid phase, against a standard sapphire reference sample. The samples were heated from 25 to 100 °C, at a rate of 10 °C/min under a nitrogen atmosphere. Importantly, all liquid photopolymer resin samples were protected from light to prevent unintended photopolymerization during preparation and measurement.

As shown in Figure S1, we found that in both liquid and solid state, $c_p$ values rise linearly with temperature. We also measured a significant decrease in $c_p$ for TJ-3704A after solidification. In our model we assume linear dependence of the heat capacity on the material state, such that

$$c_p(\Phi, T) = c_p(\Phi = 0, T) + \left(c_p(\Phi = 1, T) - c_p(\Phi = 0, T)\right) \Phi, \tag{S3}$$

where the functions

$$c_p(0, T) = c_p(0, T = 0) + \frac{\partial c_p(0, T)}{\partial T} T \tag{S4a}$$

and

$$c_p(1, T) = c_p(1, T = 0) + \frac{\partial c_p(1, T)}{\partial T} T, \tag{S4b}$$

are defined by linear fits of the measured data as shown in the figure. The resulting parameter values for all characterized materials are listed in Table S2 below.

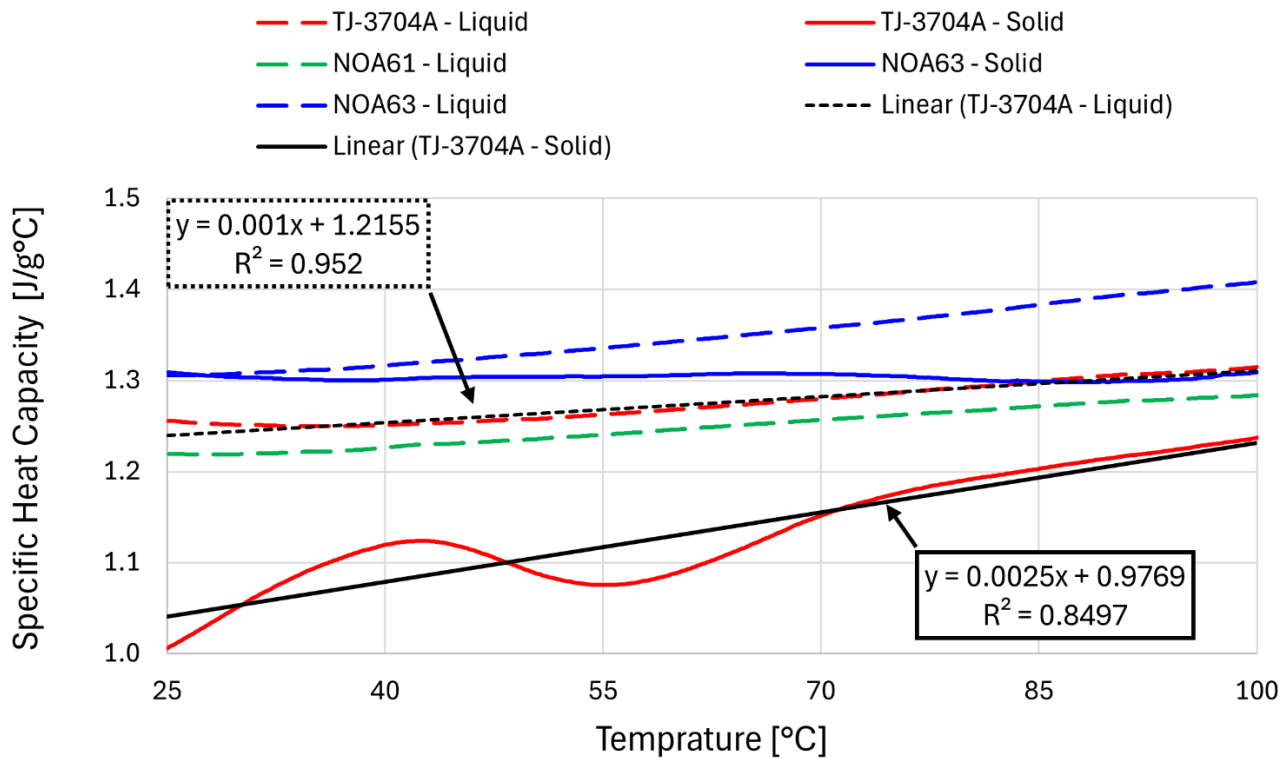

**Figure S1: Polymer specific heat capacities as measured by DSC.** We measured the polymer's specific heat capacity, $c_p$, in both liquid and solid phase from 25 to 100 °C at a rate of 10 °C/min, in a nitrogen atmosphere against a standard sapphire sample. The $c_p$ dependance on temperature within the measured range was found to be linear. We measured a significant decrease in $c_p$ for TJ-3704A (red) after solidification (solid red line in comparison to dashed red line). Based on these measurements we model the polymer's temperature dependent $c_p$ for the liquid ($\Phi = 0$) and solid ($\Phi = 1$) phases as a linear function whose coefficients are prescribed by a fit of the measured data. The black lines in the plot present an example of such a fit for the liquid TJ-3704A (dashed black) and solid TJ-3704A (solid black). As shown in eq S3, by also assuming a linear dependence of the heat capacity on the material state we extend the use of these measurements to all intermediary polymerization states.

### S4. Polymerization heat generation, $\dot{q}$, characterization by UV coupled Differential Scanning Calorimetry (UV-DSC)

Ultraviolet Differential Scanning Calorimetry (UV-DSC) enables quantitative measurement of the heat released during photopolymerization by combining conventional DSC heat-flow detection with in-situ UV irradiation. In this technique, the photopolymer resin is exposed to a controlled UV light intensity while the DSC records the exothermic heat flow associated with polymerization under a nitrogen atmosphere. By integrating the heat-flow signal over time, the total reaction enthalpy can be determined.[3]

In this work, we characterized the heat generation, $\dot{q}$, as a function of the polymerization state $\Phi$ and UV intensity $I$, for each of our polymers using a Mettler Toledo DSC-1 coupled with a Hamamatsu LC8-02 UV light source and a photocolorimetry kit. These measurements revealed typical non-monotonic autocatalytic behavior, where the reaction rate initially accelerates, reaches a peak, and then decelerates. We modeled this behavior with our autocatalytic model,

$$\dot{q} = \rho \Delta H \frac{\partial \Phi}{\partial t} = \dot{q}_0 \left(\frac{I}{I_{ref}}\right)^b \Phi^m (1-\Phi)^n, \tag{S5}$$

where $b$, $m$, $n$ and $\dot{q}_0$ are coefficients whose values we determine by data fitting, and $I_{ref}$ is an arbitrary reference value for the intensity.

**Figure S2** presents a surface plot of the autocatalytic model (heat release per unit mass as a function of polymerization state and light intensity) for TJ-3704A, overlayed by the measurements (black dots) that were used to determine its fitting parameters. The measured dependance on both UV intensity and polymerization is clearly well captured by our model. Furthermore, we note that for all three characterized polymers, such fitting resulted in typical $R^2$ values of 0.97 with root mean sum errors (RMSE) of less than 0.5 W/g. The resulting parameter values are listed in **Table S2** below.

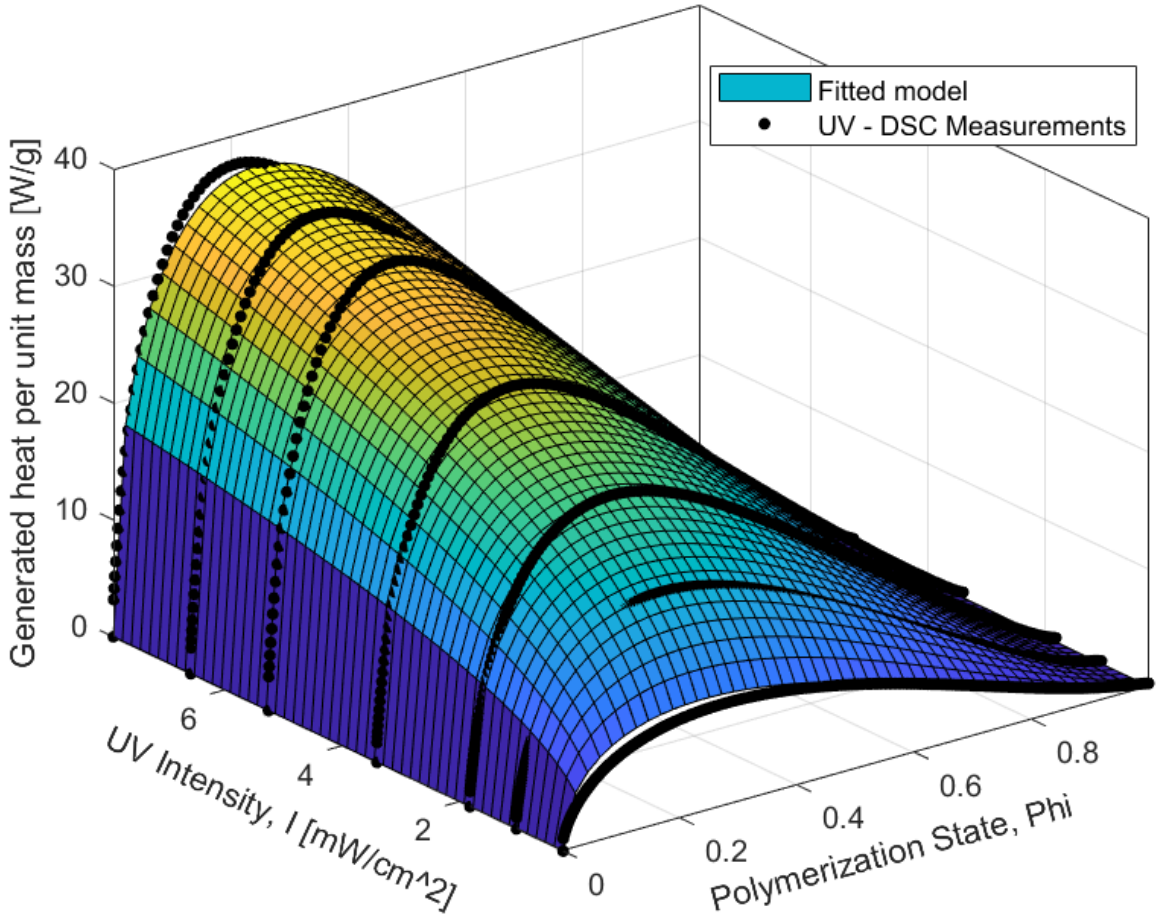

**Figure S2: Heat generation model fitting to UV-DSC measurements (black dots) of TJ-3704A.** The presented surface plot displays our model for the heat generation of TJ-3704A (normalized by its density, $\rho$) with parameters determined by fitting it to the measured data (black dots). We characterized all three photopolymers in the same manner. Fitting resulted in $R^2$ values of at least 0.97, with RMSEs lower than 0.5 W/g.

## S5. Summary of material parameters

**Table S2: Summary of material parameters**

| Property | Parameter | TJ-3704A | NOA61 | NOA63 | Units | Source |
|---|---|---|---|---|---|---|
| Density, $\rho$ | $\rho$ | 1.079 | 1.260 | ~1.260 | $\frac{g}{cm^3}$ | Polymer MSDS |
| Specific heat capacity, $c_p$ | $c_p(\Phi=0, T=0)$ | 0.9739 | 1.207 | 1.284 | $\frac{J}{gK}$ | DSC |
| | $c_p(\Phi=1, T=0)$ | 1.205 | ~1.207 | 1.292 | | |
| | $\frac{\partial c_p}{\partial T}(\Phi=0)$ | 0.0025 | 0.0007 | 0.0011 | $\frac{J}{gK^2}$ | |
| | $\frac{\partial c_p}{\partial T}(\Phi=1)$ | 0.0011 | ~0.0007 | 0.0003 | | |
| Thermal diffusivity, $\alpha$ | $\alpha_0$ | 0.11 | 0.09 | 0.14 | $\frac{mm^2}{s}$ | LFA |
| Polymer heat release, $\dot{q}$ | $\dot{q}_0$ | 57.82 | 69.53 | 90.43 | $\frac{W}{cm^3}$ | UV - DSC |
| | $b$ | 0.74 | 0.677 | 0.637 | - | |
| | $m$ | 0.53 | 0.46 | 0.39 | - | |
| | $n$ | 1.61 | 1.49 | 1.49 | - | |
| Total polymer released heat, $\Delta H$ | $H_\infty$ | 347 | 271 | 315 | $\frac{J}{g}$ | |
| | $H_0$ | 90 | 20 | 53 | | |
| | $I_{sat}$ | 13.77 | 38.25 | 38.53 | $\frac{\mu W}{cm^2}$ | |
| UV Absorption, $\beta$ | $\beta(\Phi=0)$ | 2.19 | ~2.19 | ~2.19 | $mm^{-1}$ | Transmission measurements |
| | $\beta(\Phi=1)$ | 1.49 | ~1.49 | ~1.49 | | |

~ Values which were assumed rather than directly measured

- Note: $I_{ref}$ was set as $1\frac{mW}{cm^2}$.